\documentclass[aps,prl,twocolumn,numerical,superscriptaddress,nofootinbib,showpacs,showkeys,longbibliography]{revtex4-1}

\usepackage{graphicx,color}
\usepackage{amsmath,amssymb,amsfonts}

\newcommand{\be}{\begin{equation}}
\newcommand{\ee}{\end{equation}}
\newcommand{\ba}{\begin{eqnarray}}
\newcommand{\ea}{\end{eqnarray}}
\newcommand{\ban}{\begin{eqnarray*}}
\newcommand{\ean}{\end{eqnarray*}}

\newcommand {\pA}	{$p$+Au}
\newcommand {\dA}	{$d$+Au}

\newcommand {\RuRu}	{Ru+Ru}
\newcommand {\ZrZr}	{Zr+Zr}
\newcommand {\AuAu}	{Au+Au}
\newcommand {\pPb}	{$p$+Pb}
\newcommand {\PbPb}	{Pb+Pb}
\newcommand {\Bvec}	{\vec{B}}

\newcommand {\dg}	{\Delta\gamma}

\def\v2{\mbox{$v_2$}}

\begin{document}

\title{Scaling Properties of the $\Delta\gamma$ Correlator: Constraints on Background and CME-Sensitive Charge Separation in Heavy-Ion Collisions}
\medskip

\author{Roy~A.~Lacey} 
\email{Roy.Lacey@stonybrook.edu}
\affiliation{Depts. of Chemistry \& Physics, Stony Brook University, Stony Brook, New York 11794, USA}

\author{Niseem~Magdy} 
\affiliation{Dept. of Physics, Texas Southern University, Houston, TX 77004, USA}

\date{\today}

\begin{abstract}
The scaling properties of the $\Delta\gamma$ correlator are used to
investigate charge separation associated with the chiral magnetic effect
(CME) in $p$+Au, $d$+Au, Ru+Ru, Zr+Zr, and Au+Au collisions at
$\sqrt{s_{\mathrm{NN}}}=200$~GeV, and in $p$+Pb and Pb+Pb collisions
at $\sqrt{s_{\mathrm{NN}}}=5.02$ and $2.76$~TeV. Anomalous Viscous
Fluid Dynamics (AVFD) calculations provide complementary benchmarks
for background and signal+background behavior. Scaling $\Delta\gamma$
by the elliptic-flow coefficient $v_2$ reduces the leading flow-coupled
background dependence, while the resulting $\Delta\gamma/v_2$ exhibits
a common approximate $1/N_{\rm ch}$ scaling behavior for $p$+Au,
$d$+Au, $p$+Pb, and Pb+Pb collisions, consistent with multiplicity
dilution of background-driven charge correlations across markedly
different collision systems and environments. In contrast, Ru+Ru,
Zr+Zr, and Au+Au collisions exhibit systematic positive deviations
from their respective $1/N_{\rm ch}$ background-scaling trends. These
scaling violations are qualitatively similar to those generated by
AVFD calculations with an input CME signal and, relative to the
experimentally constrained background scaling, provide evidence for a
CME contribution to the measured charge separation. The inferred CME
fraction is approximately $27\%$ for mid-central Au+Au collisions and
roughly a factor of two smaller for Ru+Ru and Zr+Zr, whose values are
consistent within uncertainties. The inferred CME fractions reduce the
expected Ru+Ru--Zr+Zr difference to only $\sim1.3\%$ in the total
charge-separation signal, explaining its limited experimental
sensitivity. These results demonstrate that scaling violations relative
to an experimentally constrained background provide a quantitative
strategy for isolating a CME contribution to charge separation in
heavy-ion collisions.
\end{abstract}

\maketitle

In relativistic ion-ion collisions, metastable domains of gluon fields
with non-trivial topological configurations~\cite{Lee:1973iz,Kharzeev:1998kz}
can form in the magnetized chiral quark-gluon plasma
(QGP)~\cite{Kharzeev:2004ey,Liao:2014ava,Miransky:2015ava,Huang:2015oca,Kharzeev:2015znc}.
The strong magnetic field, $\Bvec$, generated by the incoming ions at early
times~\cite{Kharzeev:2007jp,Asakawa:2010bu}, can couple to an imbalance
between left- and right-handed quarks induced by topological gluon-field
fluctuations, generating an electric current
$\vec{J}_V=(N_c e\vec{B}/2\pi^2)\mu_A$ along the magnetic-field
direction~\cite{Fukushima:2008xe,Son:2009tf,Zakharov:2012vv,Fukushima:2012vr}.
Here, $N_c$ is the color factor and $\mu_A$ is the axial chemical potential
characterizing the axial charge imbalance. The resulting charge separation,
known as the chiral magnetic effect (CME)~\cite{Kharzeev:2004ey}, provides
a potential experimental signature of topological QCD dynamics, but its
characterization is complicated by substantial charge-dependent background
correlations.

CME-induced charge separation is encoded in the first $P$-odd sine
coefficient, $a_1$, of the charged-particle azimuthal
distribution~\cite{Voloshin:2004vk},
\begin{eqnarray}\label{eq:a1}
\frac{dN_{\rm ch}}{d\phi}
&\propto&
1+2\sum_n
\left[
v_n\cos(n\Delta\phi)
+
a_n\sin(n\Delta\phi)
+\cdots
\right],
\end{eqnarray}
where $\Delta\phi=\phi-\Psi_{\rm RP}$ is the particle azimuthal angle
relative to the reaction-plane angle $\Psi_{\rm RP}$, and $v_n$ and $a_n$
denote the $P$-even and $P$-odd Fourier coefficients, respectively.
Because the sign of the topological charge fluctuates event by event,
$\langle a_1\rangle$ vanishes and a direct measurement of $a_1$ is not
possible. Its fluctuations, characterized by
$\tilde{a}_1=\langle a_1^2\rangle^{1/2}$, can nevertheless be accessed
through charge-sensitive observables, most notably the $\gamma$
correlator~\cite{Voloshin:2004vk}.

The $\gamma$ correlator is defined as
\begin{equation}
\gamma_{\alpha\beta} =
\left\langle
\cos\big(\phi_\alpha+\phi_\beta-2\Psi_2\big)
\right\rangle,
\qquad
\Delta\gamma =
\gamma_{\rm OS}-\gamma_{\rm SS},
\label{eq:2}
\end{equation}
where $\Psi_2$ is the second-order event-plane angle, $\phi$ is the
particle azimuthal angle, $\alpha,\beta$ denote the particle charges
$(+)$ or $(-)$, and OS and SS denote opposite-sign and same-sign
charge pairs, respectively. Experimentally, the correlator can
equivalently be evaluated using the three-particle correlation
method~\cite{Voloshin:2004vk,Abelev:2009ad,Abelev:2009ac},
\begin{equation}
\gamma_{\alpha\beta}
=
\frac{\left\langle
\cos(\phi_\alpha+\phi_\beta-2\phi_{\kappa})
\right\rangle}{v_{2,\kappa}},
\qquad
\Delta\gamma=\gamma_{\rm OS}-\gamma_{\rm SS},
\label{eq:3}
\end{equation}
where $\phi_{\kappa}$ is the azimuthal angle of a third particle
$\kappa$ used as a reference for $\Psi_2$, and $v_{2,\kappa}$ is its
elliptic-flow coefficient.

Experimental measurements indicate significant $\dg$ in \pA, \dA, \RuRu,
\ZrZr, and \AuAu\ collisions at RHIC~\cite{Abelev:2009ad,Abelev:2009ac,
Adamczyk:2013hsi,STAR:2013ksd,Adamczyk:2014mzf,STAR:2019xzd,STAR:2021mii},
and in \pPb\ and \PbPb\ collisions at the
LHC~\cite{Abelev:2012pa,ALICE:2017sss,CMS:2016wfo,CMS:2017lrw,
ALICE:2020siw}. In addition to a possible CME contribution, however,
the measured $\dg$ contains substantial charge-dependent background
correlations arising from local charge conservation, resonance decays,
and other non-flow effects~\cite{Voloshin:2004vk,Wang:2009kd,Bzdak:2009fc,
Schlichting:2010qia,Adamczyk:2013kcb,Wang:2016iov}. Many of these
background contributions are coupled to the elliptic anisotropy and
therefore scale approximately with $v_2$. The ratio $\dg/v_2$ provides
a first scaling step that largely removes this leading flow dependence,
allowing the residual charge-dependent correlations to be examined more
directly. Because many such correlations involve a finite subset of
particles embedded within a much larger event, their contribution to
$\dg/v_2$ is further expected to undergo multiplicity dilution, leading
to an approximate $1/N_{\rm ch}$ dependence. Thus, the combined $v_2$
and multiplicity scaling provides a direct test of the background-dominated
expectation without requiring the contributing background mechanisms to
be identical across collision systems. Possible departures from this
baseline can be quantified with the generalized empirical form
\begin{equation}
\frac{\Delta\gamma}{v_2}
=
a+\frac{b}{N_{\rm ch}^{\,1-c}},
\label{eq:4}
\end{equation}
where $a$ is an effective offset, $b$ sets the magnitude of the
multiplicity-dependent contribution, and $c$ quantifies the degree of
scaling violation. For background-dominated charge separation,
$c\approx0$ recovers the expected near-$1/N_{\rm ch}$ behavior, whereas
$c>0$ signals a departure from multiplicity-dilution scaling and therefore
an additional charge-separation component beyond that baseline.

An important experimental control is provided by $p(d)$+A collisions,
where CME-driven charge separation is expected to be strongly suppressed
because the magnetic field is reduced and the event plane is only weakly
correlated with the impact parameter and magnetic-field
direction~\cite{Khachatryan:2016got,Belmont:2016oqp,Kharzeev:2017uym},
while substantial charge-dependent backgrounds remain. The scaling of
$\dg/v_2$ in these systems therefore provides a direct experimental test
of the expected multiplicity-dilution behavior of background-driven
charge correlations. Its persistence in larger collision systems would
provide a stringent test of the same scaling baseline across markedly
different collision geometries, multiplicities, and particle-production
environments.

Complementing this experimental control, the AVFD
model~\cite{Shi:2017cpu,Jiang:2016wve} provides a benchmark for the
scaling patterns expected from background and signal+background
contributions to $\dg/v_2$ in A+A collisions. Comparison of these
model patterns with the experimental scaling behavior provides a basis
for assessing whether departures from the background-scaling baseline
are consistent with an additional CME-sensitive charge-separation
component.

The AVFD model~\cite{Jiang:2016wve,Shi:2017cpu} provides a complementary
benchmark for establishing the scaling behavior expected for background
and signal+background contributions to $\dg/v_2$. The event-by-event
implementation employs Monte Carlo Glauber initial conditions to evolve
fermion currents in the QGP coupled to the bulk-medium evolution described
by the VISHNU hydrodynamic framework~\cite{Shen:2014vra}, followed by a
hadronic transport stage modeled with URQMD. Charge-dependent backgrounds
arise primarily from local charge conservation (LCC) on the freeze-out
hypersurface and resonance decays. The CME contribution is generated by
the combined action of a time-dependent magnetic field,
$B(\tau)=B_0/[1+(\tau/\tau_B)^2]$, and a nonzero initial axial charge
density $n_5/s$, which induces an electric current along the magnetic-field
direction. The peak magnetic-field values $B_0$ are obtained from
event-by-event simulations~\cite{Bloczynski:2012en}, and a conservative
lifetime $\tau_B=0.6$~fm/$c$ is employed. The input $n_5/s$, associated
with topological gluonic fluctuations in the early-time glasma, and the
LCC fraction control the relative CME and background contributions.

AVFD events spanning a broad range of centralities and input values of
$n_5/s$ and LCC were used to determine the scaling patterns of
background-only and signal+background contributions to $\dg/v_2$.
The resulting comparison provides a direct model test of whether a CME
component produces a characteristic violation of the multiplicity-dilution
scaling established for the background.

%
%
\begin{figure}[t]
\includegraphics[width=1.0\linewidth, angle=-00]{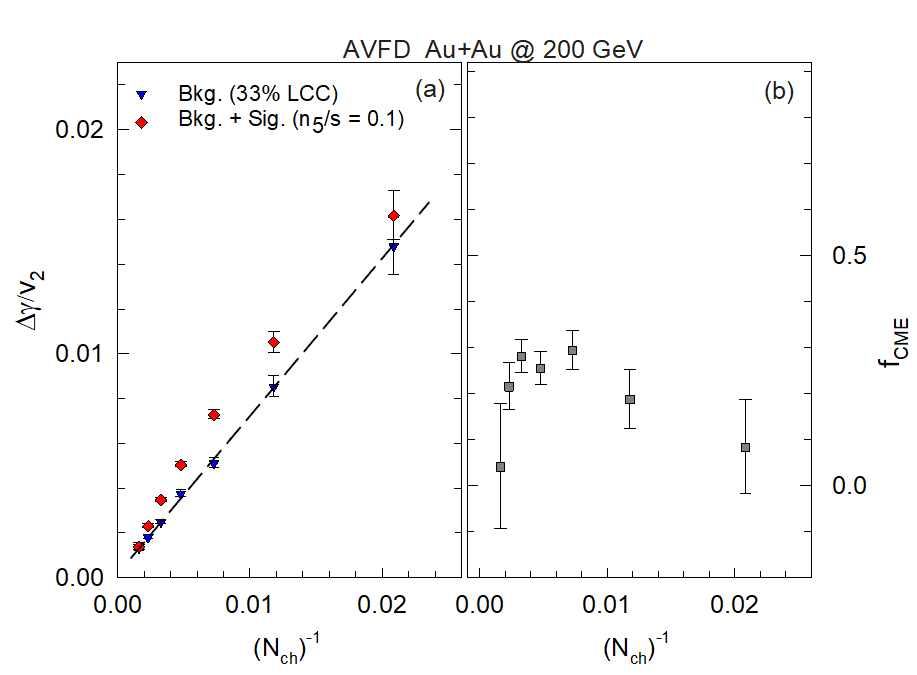}
\vskip -0.10in
\caption{
$\dg/v_2$ versus $1/N_{\rm ch}$ (a) and $f_{\rm CME}$ versus
$1/N_{\rm ch}$ (b) for simulated Au+Au collisions at
$\sqrt{s_{\mathrm{NN}}}=200$~GeV obtained with AVFD. The background-only
contribution in (a) follows the expected $1/N_{\rm ch}$ scaling, illustrated
by the dashed linear fit, whereas the signal+background contribution
deviates from this multiplicity-dilution baseline. The corresponding
$f_{\rm CME}$ values in (b), obtained from Eq.~(\ref{eq:5}), quantify the
CME-sensitive fraction of the total charge-separation signal.
}
\label{fig1}
\end{figure}

%
%
\begin{figure*}[t]
\includegraphics[width=0.75\linewidth, angle=-00]{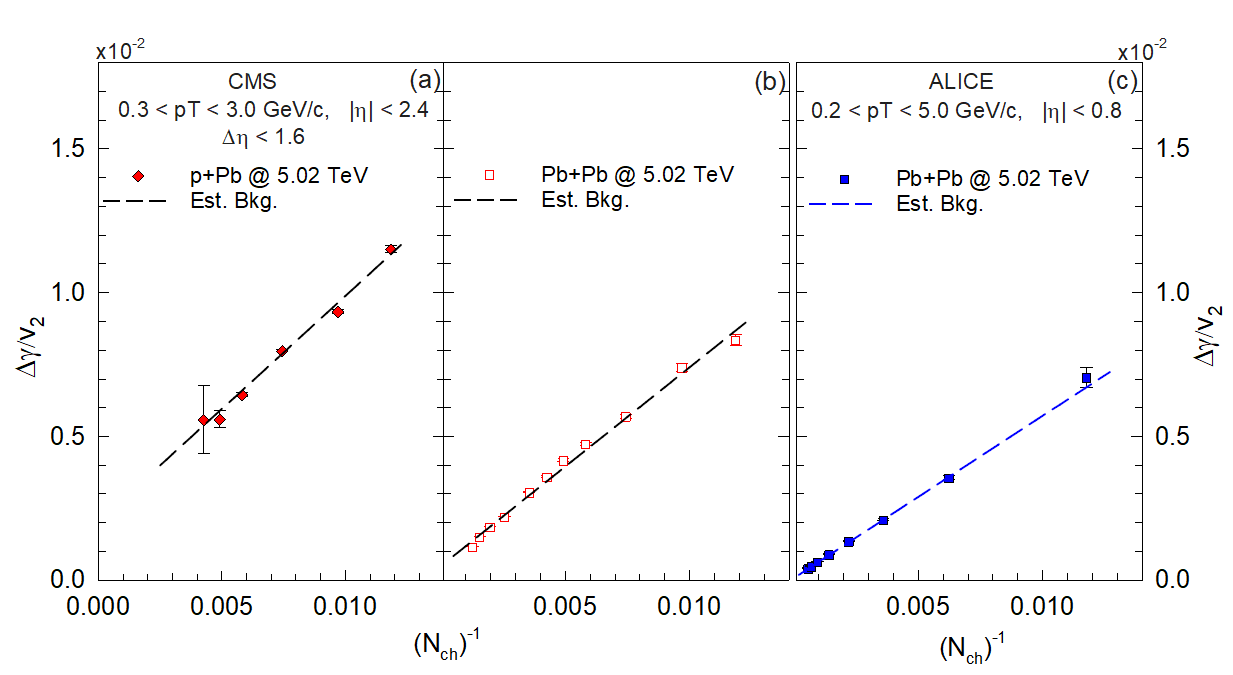}
\vskip -0.10in
\caption{
$\dg/v_2$ versus $1/N_{\rm ch}$ for \pPb\ (a) and \PbPb\ [(b) and (c)]
collisions at $\sqrt{s_{\mathrm{NN}}}=5.02$~TeV. The dashed lines indicate
the expected $1/N_{\rm ch}$ scaling associated with multiplicity dilution
of charge-dependent background correlations and provide estimates of the
background contribution. The measurements in both \pPb\ and \PbPb\
remain consistent with this scaling over the measured multiplicity range,
with no significant scaling violation observed. The data are taken from
Refs.~\cite{CMS:2016wfo,CMS:2017lrw,ALICE:2020siw}.
}
\label{fig2}
\end{figure*}

Figure~\ref{fig1}(a) demonstrates a clear separation between the AVFD
background-only and signal+background scaling patterns. The background
contribution follows the expected $1/N_{\rm ch}$ dependence associated
with multiplicity dilution, corresponding to $c\approx0$ in
Eq.~(\ref{eq:4}). In contrast, the signal+background contribution exhibits
a systematic positive departure from this baseline, corresponding to
$c>0$. Since the two calculations differ by the input CME contribution,
this result establishes, within AVFD, scaling violation as a characteristic
response of $\dg/v_2$ to CME-driven charge separation.

The magnitude of this CME-sensitive component is quantified by
\begin{equation}
f_{\rm CME}
=
\frac{
[\dg/v_2(\mathrm{Sig.}+\mathrm{Bkg.})
-
\dg/v_2(\mathrm{Bkg.})]
}
{
[\dg/v_2(\mathrm{Sig.}+\mathrm{Bkg.})]
},
\label{eq:5}
\end{equation}
shown as a function of $1/N_{\rm ch}$ in Fig.~\ref{fig1}(b). For the
illustrative inputs $n_5/s=0.1$ and LCC$=33\%$, the calculation gives
$f_{\rm CME}\approx30\%$ for 30--40\% central Au+Au collisions.
The CME-sensitive fraction peaks in mid-central collisions and approaches
zero toward both central and peripheral collisions, where the background
and signal+background values of $\dg/v_2$ become comparable. This behavior
motivates the use of the central and peripheral measurements as anchors
for constraining the background contribution across the full centrality
range.

%
%
\begin{figure*}[t]
\includegraphics[width=0.75\linewidth, angle=-00]{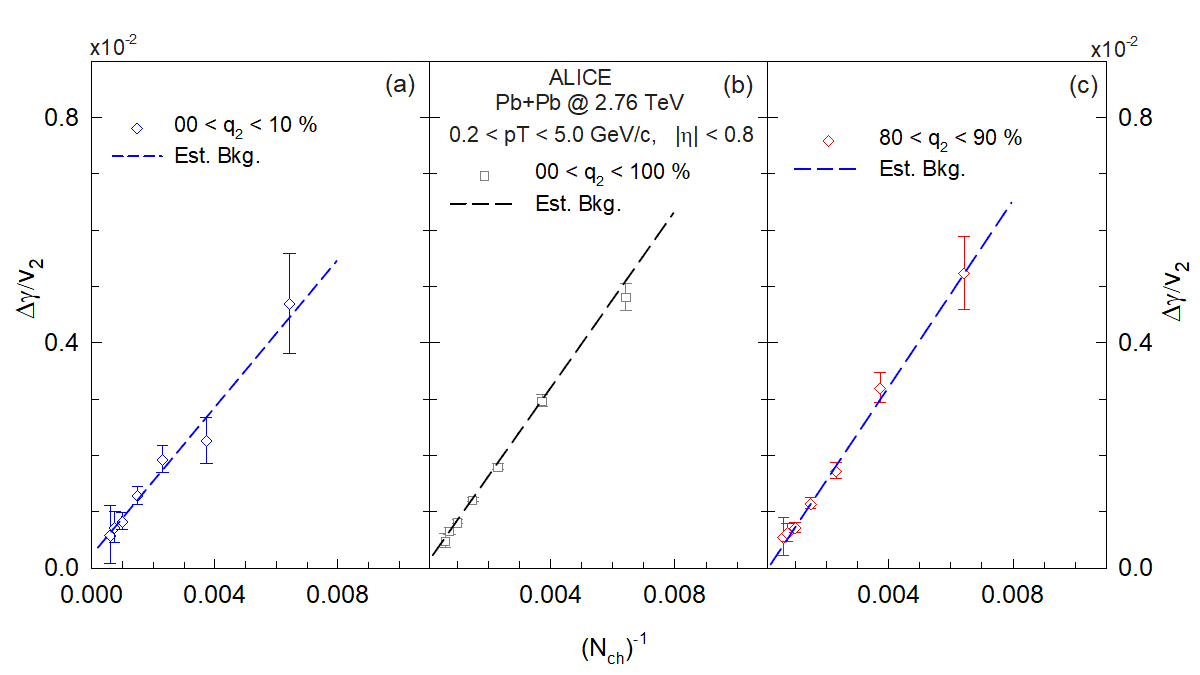}
\vskip -0.10in
\caption{
$\dg/v_2$ versus $1/N_{\rm ch}$ for $q_2$-selected \PbPb\ collisions at
$\sqrt{s_{\mathrm{NN}}}=2.76$~TeV. The dashed lines indicate the expected
$1/N_{\rm ch}$ scaling associated with multiplicity dilution of
charge-dependent background correlations and provide estimates of the
background contribution. The measurements remain consistent with the
same scaling trend across the different event-shape selections. The data
are taken from Ref.~\cite{ALICE:2017sss}.
}
\label{fig3}
\end{figure*}
%
%

%
%
\begin{figure*}[tb]
\includegraphics[width=0.75\linewidth, angle=-00]{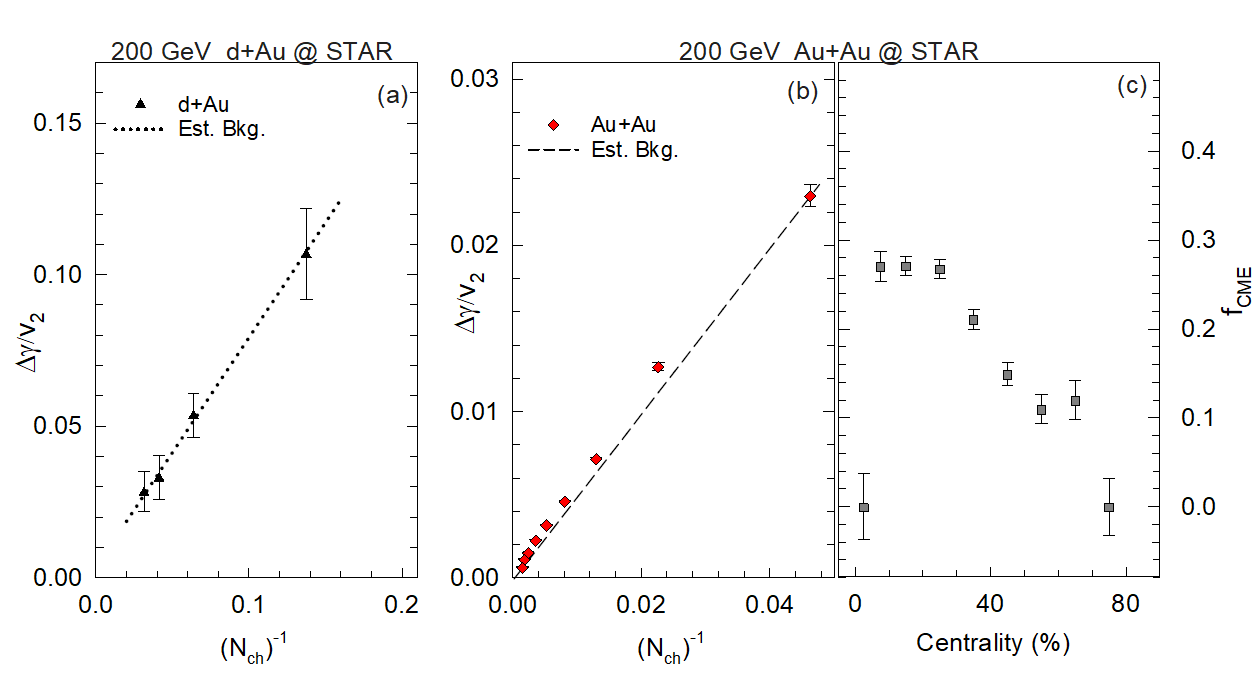}
\vskip -0.10in
\caption{
$\dg/v_2$ versus $1/N_{\rm ch}$ [(a) and (b)] and the extracted
$f_{\rm CME}$ versus centrality (c) for \dA\ and \AuAu\ collisions at
$\sqrt{s_{\mathrm{NN}}}=200$~GeV. Panel (a) provides an experimental
benchmark for the expected background scaling. The dotted and dashed
lines indicate the expected $1/N_{\rm ch}$ scaling associated with
multiplicity dilution of charge-dependent background correlations and
provide estimates of the background contribution. The $f_{\rm CME}$
values in panel (c), obtained from Eq.~(\ref{eq:5}), quantify the
CME-sensitive fraction extracted for \AuAu\ collisions as a function
of centrality.
}
\label{fig4}
\end{figure*}
%
%
%
\begin{figure*}
\centering
\includegraphics[width=0.49\textwidth]{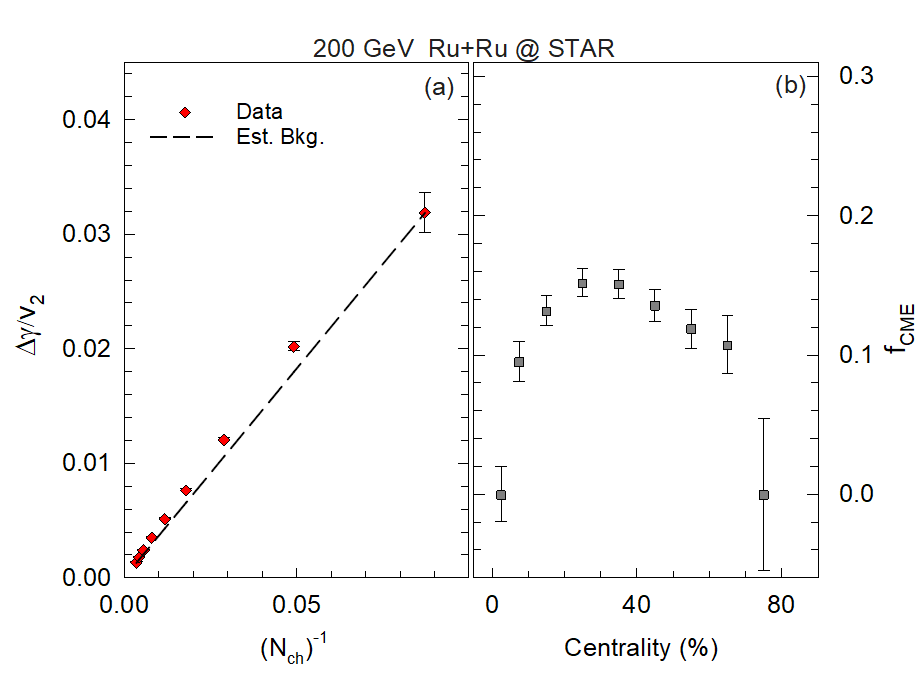}
\includegraphics[width=0.49\textwidth]{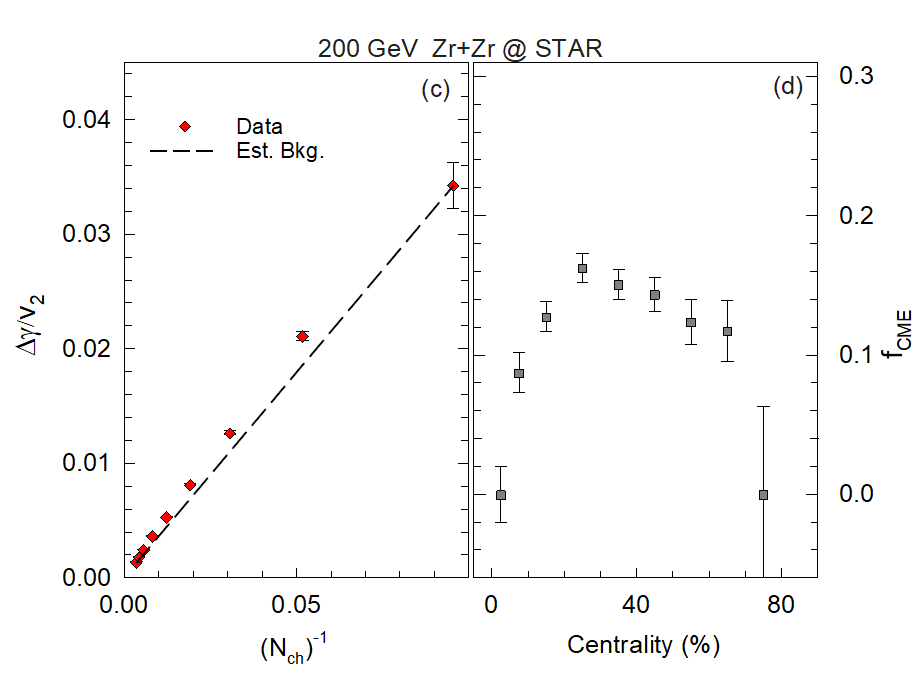}
\caption{
$\dg/v_2$ versus $1/N_{\rm ch}$ [(a) and (c)] and $f_{\rm CME}$ versus
centrality [(b) and (d)] for \RuRu\ and \ZrZr\ collisions at
$\sqrt{s_{\mathrm{NN}}}=200$~GeV. The dashed lines indicate estimates
of the background contributions. The $f_{\rm CME}$ values in (b) and
(d), obtained from Eq.~(\ref{eq:5}), quantify the extracted CME-sensitive
fraction of the charge separation. The data are taken from
Ref.~\cite{STAR:2021mii}.
}
\label{fig5}
\end{figure*}

The small $f_{\rm CME}$ values predicted by AVFD toward central and
peripheral collisions provide a physical basis for using these regions
as anchors for the background scaling. In central collisions, the CME
contribution is suppressed by the reduced magnetic field, whereas in
peripheral collisions it is suppressed by the substantial decorrelation
between the event plane and the magnetic-field direction. Importantly,
the anchor measurements do not define the background independently.
Rather, they constrain its normalization in conjunction with the
$1/N_{\rm ch}$ dependence motivated by multiplicity dilution and
independently constrained by measurements in background-dominated
collision systems. Together, the experimentally constrained
$1/N_{\rm ch}$ scaling and the central and peripheral anchors therefore
define a data-constrained background baseline across the intervening
centrality range. Deviations from this baseline can then be used to
quantify an additional charge-separation component.

This construction is also conservative with respect to possible residual
CME-sensitive contributions in the anchor regions. Such contributions
would raise the inferred background level and thereby reduce, rather than
enhance, the extracted excess. The resulting CME-sensitive fraction would
therefore be underestimated rather than artificially amplified.

The $v_2$ and $\dg$ measurements for \pA, \dA, \RuRu, \ZrZr, and
\AuAu\ collisions at RHIC~\cite{Abelev:2009ad,Abelev:2009ac,
Adamczyk:2013hsi,Adamczyk:2014mzf,STAR:2019xzd,STAR:2021mii},
together with \pPb\ and \PbPb\ collisions at the
LHC~\cite{ALICE:2017sss,CMS:2016wfo,CMS:2017lrw,ALICE:2020siw},
provide a broad experimental basis for testing the scaling properties
of $\dg/v_2$ across markedly different collision systems, energies,
geometries, and multiplicities. Figure~\ref{fig2} shows the results for
\pPb\ and \PbPb\ collisions at $\sqrt{s_{\mathrm{NN}}}=5.02$~TeV.
The measurements exhibit a common approximate $1/N_{\rm ch}$ dependence
over a broad multiplicity range, consistent with the expected
multiplicity dilution of background-driven charge correlations. The
\pPb\ measurements provide a particularly important experimental control
because CME-driven charge separation is expected to be strongly
suppressed in these collisions. Their observed scaling therefore
demonstrates that the substantial charge-dependent backgrounds present
in \pPb\ follow the expected multiplicity-dilution behavior without
generating a significant scaling violation.

The persistence of the same scaling behavior in \PbPb\ collisions,
despite the markedly different collision geometry, multiplicity, and
particle-production environment, provides a stringent cross-system test
of the background-scaling baseline and indicates that any additional
CME-sensitive contribution is small relative to the dominant
background-driven correlations. An independent stress test is provided
by the $q_2$-selected \PbPb\ measurements at
$\sqrt{s_{\mathrm{NN}}}=2.76$~TeV shown in Fig.~\ref{fig3}
~\cite{ALICE:2017sss}. Event-shape selection based on fractional cuts
of the $q_2$ flow-vector distribution~\cite{Schukraft:2012ah} produces
event classes with substantially different elliptic-flow magnitudes and
underlying geometries. Nevertheless, the corresponding $\dg/v_2$
measurements remain consistent with the same approximate $1/N_{\rm ch}$
dependence. The survival of the scaling across both collision systems
and event-shape selections provides strong experimental support for the
robustness of the multiplicity-dilution baseline for background-driven
charge correlations.

The scaling results for RHIC collisions at
$\sqrt{s_{\mathrm{NN}}}=200$~GeV are shown in
Figs.~\ref{fig4} and \ref{fig5}. The \dA\ measurements exhibit the same
approximate $1/N_{\rm ch}$ dependence observed for \pPb\ collisions
[cf. Fig.~\ref{fig2}(a)], consistent with the expectation that
CME-driven charge separation is strongly suppressed in $p(d)$+A
collisions. Similar behavior is observed for \pA\ collisions (not shown),
albeit with larger statistical uncertainties. Together with the \pPb\
and \PbPb\ results, these measurements establish a common
$1/N_{\rm ch}$ scaling behavior for background-driven charge separation
across markedly different collision systems, energies, geometries, and
particle-production environments. This common scaling does not imply
identical background magnitudes or microscopic compositions across
systems; rather, it reflects the shared multiplicity-dilution behavior
of the dominant charge-dependent correlations.

Against this experimentally constrained background-scaling behavior,
the larger RHIC systems exhibit a qualitatively different pattern. The
\AuAu\ [Fig.~\ref{fig4}(b)], \RuRu\ [Fig.~\ref{fig5}(a)], and \ZrZr\
[Fig.~\ref{fig5}(c)] measurements show systematic positive deviations
from their respective $1/N_{\rm ch}$ background-scaling trends. These
scaling violations are qualitatively similar to those generated by the
signal+background AVFD calculations in Fig.~\ref{fig1}(a) and, relative
to the experimentally constrained background scaling, provide evidence
for a CME contribution to the measured charge separation.

To quantify this contribution, $f_{\rm CME}$ [Eq.~(\ref{eq:5})] is
evaluated relative to the background baseline constrained by the
experimentally observed $1/N_{\rm ch}$ scaling and the central and
peripheral anchors described above. The resulting background estimates
are indicated by the dashed lines in Figs.~\ref{fig4}(b),
\ref{fig5}(a), and \ref{fig5}(c), and the corresponding $f_{\rm CME}$
values are shown in Figs.~\ref{fig4}(c), \ref{fig5}(b), and
\ref{fig5}(d). The inferred CME fractions exhibit a clear centrality
dependence and indicate a non-negligible CME contribution. For
mid-central \AuAu\ collisions, $f_{\rm CME}\sim27\%$, approximately
twice the values obtained for \RuRu\ and \ZrZr.

Within the quoted uncertainties, no significant difference is observed
between the inferred $f_{\rm CME}$ values for \RuRu\ and \ZrZr. This
does not, however, imply comparable underlying CME signal strengths.
For an expected $\sim10\%$ difference in the CME signal between the two
isobars~\cite{STAR:2021mii}, an inferred CME fraction of $\sim13\%$
corresponds to only a $\sim1.3\%$ difference in the total measured
charge-separation correlator. Such a difference lies well below the
current experimental uncertainties. Thus, the absence of a resolved
Ru+Ru--Zr+Zr difference is compatible with the inferred CME fractions
while underscoring the limited sensitivity of the present $\dg$
measurements to the expected isobar signal difference.

In summary, the scaling properties of the $\dg$ correlator provide new
experimental constraints on the origin of charge separation across a
broad range of collision systems at RHIC and the LHC. Scaling $\dg$ by
$v_2$ largely removes the leading elliptic-flow dependence of the
background, while the remaining $\dg/v_2$ correlations in \pA\ and
\dA\ collisions at $\sqrt{s_{\mathrm{NN}}}=200$~GeV, \pPb\ collisions
at $\sqrt{s_{\mathrm{NN}}}=5.02$~TeV, and \PbPb\ collisions at
$\sqrt{s_{\mathrm{NN}}}=5.02$ and $2.76$~TeV exhibit a common
approximate $1/N_{\rm ch}$ dependence consistent with multiplicity
dilution of background-driven charge correlations. The persistence of
this scaling across markedly different geometries, multiplicities,
particle-production environments, and background compositions establishes
a robust experimental background-scaling behavior without requiring the
background magnitudes or microscopic sources to be identical across
systems. Against this experimentally constrained background-scaling
behavior, the \AuAu, \RuRu, and \ZrZr\ measurements exhibit systematic
positive deviations from their respective $1/N_{\rm ch}$ trends. These
scaling violations are qualitatively similar to those generated by AVFD
calculations with an input CME signal and, relative to the experimentally
constrained background scaling, provide evidence for a CME contribution
to the measured charge separation. Combining the experimentally
constrained $1/N_{\rm ch}$ dependence with central and peripheral anchor
measurements provides system-specific background baselines from which
this contribution can be quantified. The inferred CME fraction is approximately 
$27\%$ for mid-central Au+Au collisions and roughly a factor of two smaller for 
Ru+Ru and Zr+Zr, whose values are consistent within uncertainties. The inferred
CME fractions reduce the expected Ru+Ru--Zr+Zr difference to only
$\sim1.3\%$ in the total charge-separation signal, explaining its
limited experimental sensitivity. These results demonstrate that
scaling violations relative to an experimentally constrained background
provide a quantitative strategy for isolating a CME contribution to
charge separation in heavy-ion collisions.

\section*{Acknowledgments}
\begin{acknowledgments}
This research is supported by the US Department of Energy, Office of Science, Office of Nuclear Physics, 
under contracts DE-SC0024602.
%
\end{acknowledgments}
%
%
\bibliography{lpvpub} 
\end{document}